\documentstyle[a4,12pt,psfig]{article}

\begin{document}
\pagestyle{empty}
\setcounter{page}{1}

\centerline{\bf \Large{CONTROL OF FRACTIONAL-ORDER CHUA'S SYSTEM}}
\vspace*{0.37truein}
\centerline{\footnotesize IVO PETRAS}
\vspace*{0.05truein}
\centerline{\footnotesize\it Department of Informatics and Process Control} 
\centerline{\footnotesize\it Technical University of Kosice}
\centerline{\footnotesize\it B.Nemcovej 3,  042 00 Kosice,
Slovak Republic }
\centerline{\footnotesize\it e-mail: petras@tuke.sk}

\vspace*{10pt}
%\vspace*{0.225truein}
\normalsize
%\vspace*{0.21truein}
\begin{abstract}
\hspace{-7.5mm}
This paper deals with feedback control of fractional-order
Chua's system. The fractional-order Chua's system with total order less
than three which exhibit chaos as well as other nonlinear behavior and
theory for control of chaotic systems using sampled data are presented.
Numerical experimental example is shown to verify the theoretical results.
\end{abstract}

\vspace*{1pt}
\section{Introduction}          
\vspace*{-0.5pt}
\noindent
It is well known that chaos cannot occur in continuous systems
of total order less than three. This assertion is based on the usual
concepts of order, such as the number of states in a system or
the total number of separate differentiations or integrations in
the system. The model of system can be rearranged to three single
differential equations, where one of the equations contains the
non-integer (fractional) order  derivative. The total order
of system is changed from $3$ to $2+q$, where $0<q\leq1$.
To put this fact into context, we can consider the fractional-order
dynamical model of the system.
Hartley {\it et al.} \cite{4} consider the fractional-order
Chua's system and demonstrated that chaos is possible for systems
where the order is less than three. In their work, the limits on
the mathematical order of the system to have a chaotic response, as
measured from  the bifurcation diagrams, are approximately from
$2.5$ to $3.8$.
In work \cite{1}, chaos was discovered in fractional-order
two-cell cellular neural networks and also in work \cite{6}
chaos was exhibited in a system with total order less than three.

The control of chaos has been
studied and observed in experiments (e.g. works \cite {3},
\cite{5}, \cite{8}, \cite{12}). Especially, the control of well-known
Chua's system \cite{9} by sampled data has been studied
\cite{13}. The main motivation for the control of chaos via sampled
data is well-developed digital control techniques.

In this brief studies are presented practical results from
sampled-data feed-back control of fractional-order chaotic
dynamical system, modeled by the state equation
$\dot {\bf x} = {\bf f(x)}$, where ${\bf x}\in \Re^n$ is state variable,
${\bf f}: \Re^n \to \Re^n$ is nonlinear function and ${\bf f}(0)=0$.

The approach used in this paper is concentrating on the feed-back
control of the chaotic fractional-order Chua's system, where
total order of the system is $2.9$.

\section{Fractional calculus}
\subsection{Definitions of Fractional Derivatives }
The idea of fractional calculus has been known since the development
of the regular calculus, with the first reference
probably being associated with Leibniz and L'Hospital in 1695.

Fractional calculus is a generalization of integration and differentiation
to non-integer order fundamental operator $_{a}D^{\alpha}_{t}$,
where $a$ and $t$ are the limits of the operation.
The continuous integro-differential operator is defined as
$$
 _aD^{\alpha}_{t} = \left \{
        \begin{array}{ll}
                \frac{d^{\alpha}}{dt^{\alpha}} & \mbox{$\Re(\alpha)>0$,} \\
                 1 & \mbox{$\Re(\alpha)=0$,} \\
                \int_{a}^{t} (d\tau)^{-\alpha} & \mbox{$\Re(\alpha)<0$.}
        \end{array}
        \right.
$$
The two definitions used for the general fractional differintegral
are the Gr\"unwald-Letnikov (GL) definition and the Riemann-Liouville (RL)
definition \cite{7}, \cite{11}. The GL is given here
\begin{equation}\label{GLD_d}
     _{a}D^{\alpha}_{t}f(t)=\lim_{h \to 0} h^{-\alpha}
     \sum_{j=0}^{[\frac{t-a}{h}]}(-1)^j {\alpha \choose j}f(t-jh),
\end{equation}
where $[x]$ means the integer part of $x$. The RL definition is given as
\begin{equation}\label{LRL}
   _{a}D_{t}^{\alpha}f(t)=
    \frac{1}{\Gamma (n -\alpha)}
    \frac{d^{n}}{dt^{n}}
    \int_{a}^{t}
    \frac{f(\tau)}{(t-\tau)^{\alpha - n + 1}}d\tau, \\
\end{equation}
for $(n-1 < \alpha <n)$ and
where $\Gamma (.)$ is the {\it Gamma}  function.

\subsection{Numerical Methods for Calculation of Fractional
\protect \\ Derivatives}
For numerical calculation of fractional-order derivation we can use the
relation (\ref{FD}) derived from the Gr\"unwald-Letnikov definition
(\ref{GLD_d}). This approach is based on the fact that a wide class
of functions, two definitions - GL (\ref{GLD_d}) and RL
(\ref{LRL}) - are equivalent. The relation
for the explicit numerical approximation of $\alpha$-th derivative at the points $kT,\, (k = 1, 2,\dots)$ has the following form
\cite{2}, \cite{10}, \cite{11}:
\begin{equation}\label{FD}
    _{(k-L/T)}D^{\alpha}_{kT} f(t) \approx
      T^{-\alpha}  \sum_{j=0}^{k} (-1)^j {\alpha \choose j} f_{k-j},
\end{equation}
where $L$ is the "memory length", $T$ is the step size of the
calculation (sample period)
and $(-1)^j {\alpha \choose j}$ are binomial coefficients $c_j^{(\alpha)},
\,(j=0, 1, \dots)$. For its calculation we can use the following expression:
\begin{equation}\label{b_k}
  c_0^{(\alpha)} = 1, \qquad c_j^{(\alpha)} = \left (1 -
                       \frac{1+ \alpha}{j}\right)c_{j-1}^{(\alpha)}.
\end{equation}

\subsection{Some Properties of Fractional Derivatives }

Two general properties of fractional derivative we will be used.
The first is composition of fractional with integer-order derivative
and the second is the property of linearity.

The fractional-order derivative commutes with integer-order derivation
\cite{11},
\begin{equation}\label{commut}
\frac{d^n}{dt^n} (_aD^p_t f(t))= \,_aD^p_t \left ( \frac{d^n f(t)}{dt^n}\right ) =\, _aD_t^{p+n}f(t),
\label{KR}
\end{equation}
under the condition $t=a$ we have
$f^{(k)}(a)=0,\,(k=0, 1, 2,\dots,n-1)$.
The relationship (\ref{KR}) says the operators $\frac{d^n}{dt^n}$ and $_aD_t^p$ commute.

Similar to integer-order differentiation, fractional differentiation is
a linear operation \cite{11}:
\begin{equation}\label{linear}
_aD^p_t \left( \lambda f(t) + \mu g (t) \right )= \lambda\,
_aD^p_t f(t) + \mu\, _aD^p_t g(t).
\end{equation}

\section{Fractional-Order Chua's System}
\noindent

Classical Chua's oscillator, which is shown in Fig. 1, is given
by
\begin{eqnarray}\label{Chua_o}
    \frac{d v_1}{dt} &=& \frac{1}{C_1} \left [G(v_2 - v_1) - f(v_1) \right], \nonumber \\
    \frac{d v_2}{dt} &=& \frac{1}{C_2} \left [G(v_1 - v_2) +i \right ],  \\
    \frac{d i  }{dt} &=& \frac{1}{L}\left [ -v_2(t) - R_L i \right] , \nonumber
\end{eqnarray}
where $G=1/R$ and $f(v_1)$ is the piecewise linear $v-i$
characteristic of nonlinear Chua's diode.

Given the techniques of fractional calculus, there are still a
number of ways in which the order of system could be amended. One
approach  would be to change the order of any or all of tree
constitutive equations (\ref{Chua_o}) so that the total
order gave the desired value.
\begin{figure}[ht]
\vspace*{13pt}
%\centerline{\vbox{\hrule width 8cm height0.001pt}}
%\centerline{\hbox to 7 cm {\special{em:graph obr_chua.bmp}}
%\hbox to 7 cm {\special{em:graph obr_va.bmp}}}
\centerline{\psfig{file=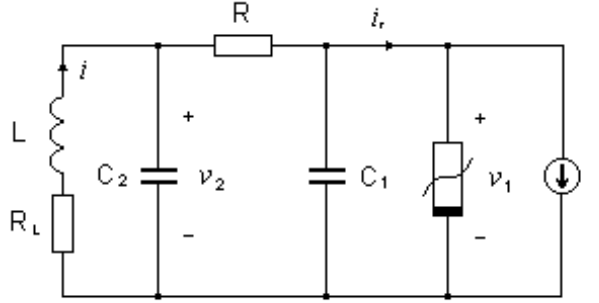,height=4.7cm}} %100 percent
%\centerline{\psfig{file=fig1_b.eps}} %100 percent
%\vspace*{4.7 cm}
%\centerline{\vbox{\hrule width 8cm height0.001pt}}
%\vspace*{13pt}
\caption{\it Chua's circuit.}
\end{figure}

In our case, in the equation one, we replace the first
differentiation by fractional differentiation of order $q$, $q \in
\Re$. The final dimensionless equations of the system for $R_L=0$
are $(x_1 = v_1, x_2 = v_2, x_3 = i)$:
\begin{eqnarray}\label{Chua-Hartley}
    \frac{d x_1(t)}{dt} &=& \alpha \,\,_0D^{1-q}_t \left( x_2(t) -
                                x_1(t) - f(x_1) \right ),
                          \nonumber \\
    \frac{d x_2(t)}{dt} &=& x_1(t) - x_2(t) +  x_3(t),  \\
    \frac{d x_3(t)}{dt} &=& -\beta x_2(t), \nonumber
\end{eqnarray}
where
$$
f(x_1)=b x_1 + \frac{1}{2}(a-b)(|x_1+1|-|x_1-1|)
$$
and $\alpha = C_2/C_1$, $\beta=C_2R^2/L$.

\section{Feedback Control of Chaos}

The structure of control system with sampled data \cite{13} is shown in
Fig. 2. The state variables of the chaotic system are measured
and the result is used to construct the output  signal ${\bf y}(t)
=D{\bf x}(t)$, where $D$ is a constant matrix. The output  ${\bf y}(t)$
is then sampled by sampling block to obtain ${\bf y}(k) = D {\bf x}(k)$
at the discrete moments $kT$, where $k=0, 1, 2, \dots$, and $T$ is the
sample period. Then $D{\bf x}(k)$ is used by the controller to calculate
the control signal ${\bf u}(k)$, which is fed back into chaotic system.

The controlled chaotic system is defined by relations \cite{13}
\begin{eqnarray}\label{CCHS}
\frac{d {\bf{x}(t)}}{dt} &=& {\bf{f}}({\bf x}(t))+ B {\bf u}(k),
\,\,\,\,\, t \in [kT, (k+1)T) \nonumber \\
{\bf u}(k+1) &=& C {\bf u}(k) + D {\bf x}(k),\,\,\,\,\, k=0, 1, 2, \dots
\end{eqnarray}
where ${\bf u}\in\Re^m,\, B \in \Re^n \times \Re ^m, \,C \in \Re^m \times \Re^m,\, D \in \Re^m \times \Re^n$ and $t\in R_+$; ${\bf x}(k)$ is the sampled value of ${\bf x}(t)$ at $t=kT$. Observe that since ${\bf f(0)=0}$ is an equilibrium point of the system (\ref{CCHS}).

The controlled fractional-order Chua's system is defined by
\begin{eqnarray}\label{CCS}
    \frac{d x_1(t)}{dt} &=& \alpha \,\,_0D^{1-q}_t \left( x_2(t) -
                                x_1(t) - f(x_1) \right ) + u_1(t),
                          \nonumber \\
    \frac{d x_2(t)}{dt} &=& x_1(t) - x_2(t) +  x_3(t) + u_2(t),  \\
    \frac{d x_3(t)}{dt} &=& -\beta x_2(t) + u_3(t). \nonumber
\end{eqnarray}

\begin{figure}[ht]
\vspace*{15pt}
%\centerline{\vbox{\hrule width 10cm height0.001pt}}
\centerline{\psfig{file=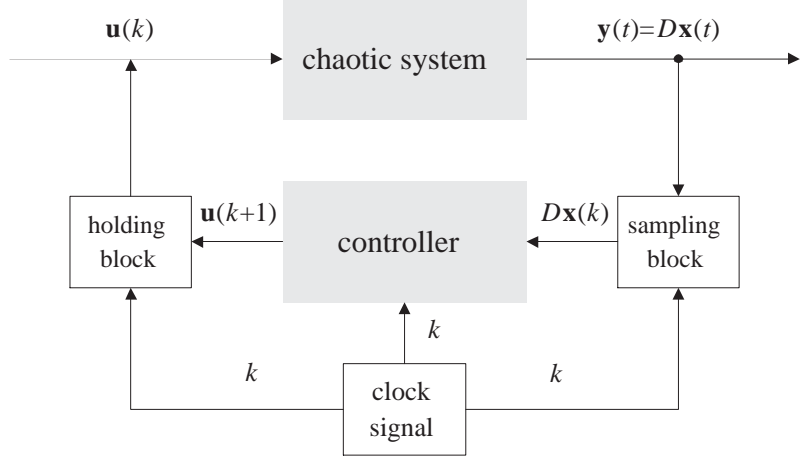,height=6.2 cm}} %100 percent
%\vspace*{5.5cm}         %ORIGINAL SIZE=1.6TRUEIN x 100% - 0.2TRUEIN
%\centerline{\vbox{\hrule width 10cm height0.001pt}}
\vspace*{13pt}
\caption{\it Structure of control system.}
\end{figure}

\section{Illustrative Example}

For numerical simulations the following parameters
of the fractional Chua's system (\ref{Chua-Hartley}) were chosen:
$$\alpha = 10,\,\,\beta=\frac{100}{7},\,\,q=0.9,\,\,
a=-1.27,\,\,b=-0.68,$$
and the following parameters (experimentally found) of controller:
\begin{equation}\label{parameters_contrl}
B=\left (
\begin{array}{ccc}
1       & 0     & 0 \\
0       & 0     & 0 \\
0       & 0     & 0 \\
\end{array}
\right ),\,\,
C=\left (
\begin{array}{ccc}
0.8     & 0     & 0 \\
0       & 0     & 0 \\
0       & 0     & 0 \\
\end{array}
\right ),\,\,
D=\left (
\begin{array}{ccc}
-3.3    & 0     & 0 \\
0       & 0     & 0 \\
0       & 0     & 0 \\
\end{array}
\right ).
\end{equation}
Using the above parameters (\ref{parameters_contrl})
the digital controller in state space form is defined as
\begin{equation}\label{DC}
    u_1(k+1) =  0.8 u_1(k) - 3.3 x_1(k),
\end{equation}
for $k=0, 1, 2,\dots$
The initial conditions for Chua's circuit were $((x_1(0), x_2(0),
x_3(0))$ = $(0.2, -0.1, -0.01)$ and the initial condition for
the controller (\ref{DC}) was $((u_1(0)$ = $(0))$.
The sampling period (frequency) was $T=100$ Hz.

For the computation of the fractional-order derivative in equations
(\ref{CCS}), the relations (\ref{FD}), (\ref{b_k})
and properties (\ref{commut}), (\ref{linear}) were used.
The length of memory was $L=10$\, ($1000$ coefficients for $T=100$ Hz).

Fig. 3 shows the attractor of Chua's circuit (\ref{Chua-Hartley})
without control.
Similar behaviour was shown in work \cite{4}, where
piecewise linear nonlinearity was replaced by cubic nonlinearity
which yields very similar behaviour.
\begin{figure}
\vspace*{13pt}
%\centerline{\vbox{\hrule width 10cm height0.001pt}}
\centerline{\psfig{file=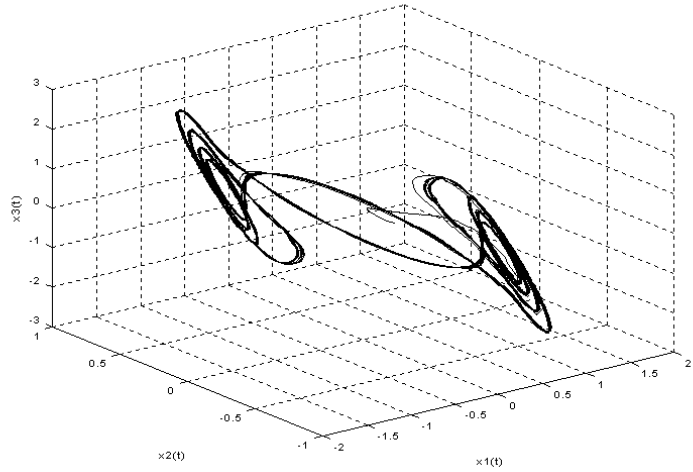,height=7cm}} %100 percent
%\vspace*{6.5cm}         %ORIGINAL SIZE=1.6TRUEIN x 100% - 0.2TRUEIN
%\centerline{\vbox{\hrule width 10cm height0.001pt}}
%\vspace*{13pt}
\caption{\it Strange attractor of fractional-order Chua's system without control.}
\end{figure}

\begin{figure}
\vspace*{13pt}
%\centerline{\vbox{\hrule width 10cm height0.001pt}}
\centerline{\psfig{file=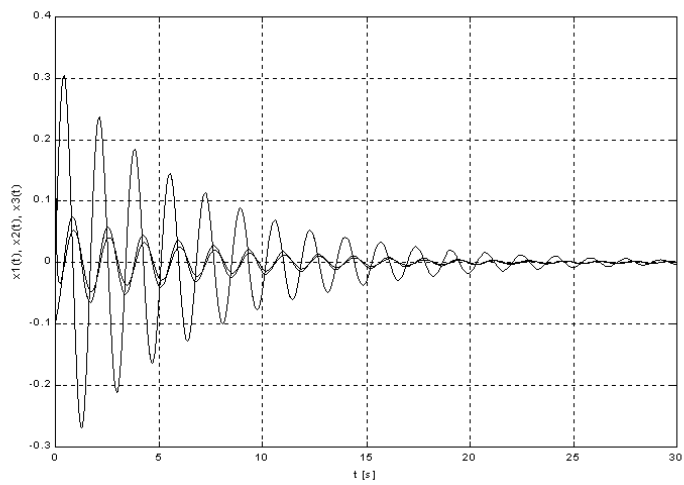,height=7cm}} %100 percent
%\vspace*{6.5cm}         %ORIGINAL SIZE=1.6TRUEIN x 100% - 0.2TRUEIN
%\centerline{\vbox{\hrule width 10cm height0.001pt}}
%\vspace*{13pt}
\caption{\it Controlled state variables $x_1(t), x_2(t), x_3(t)$.}
\end{figure}

\begin{figure}
\vspace*{13pt}
%\centerline{\vbox{\hrule width 10cm height0.001pt}}
\centerline{\psfig{file=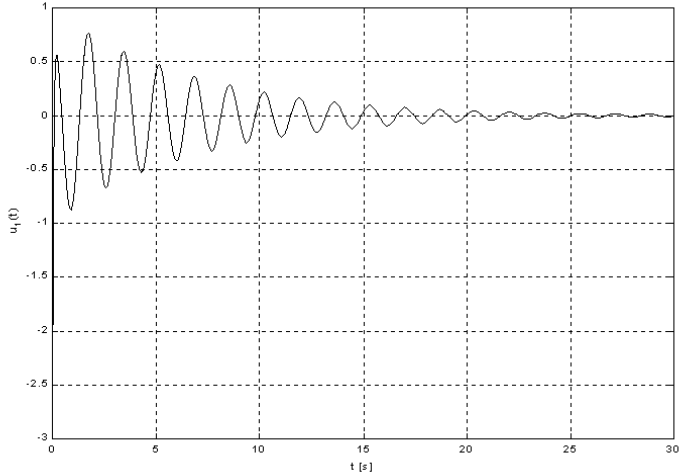,height=7cm}} %100 percent
%\vspace*{6cm}         %ORIGINAL SIZE=1.6TRUEIN x 100% - 0.2TRUEIN
%\centerline{\vbox{\hrule width 10cm height0.001pt}}
%\vspace*{13pt}
\caption{\it Control signal $u_1(t)$.}
\end{figure}

In Fig. 4 is shown the controlled trajectory of state
variables of the fractional-order Chua's system (\ref{CCS}),
which tends to origin asymptotically.

In Fig. 5 is shown control signal from the digital controller
(\ref{DC}).

\section{Conclusion}

We have considered an example of control of chaotic
fractional-order Chua's circuit, which exhibits chaotic behaviour
with total order less than three. As has been demonstrated,
the idea of fractional calculus requires one to reconsider
dynamic system concepts that are often taken for granted. So by
decreasing the order of a system from $3$ to $2+q$ in this way,
we also move from a three-dimensional system to one of infinite
dimension. This system can be controlled by sampled data. The
sampled data of output are sufficient for constructing  the control
signals in the digital controller. Digital controllers had been
widely used in industry.

The conclusion of this work confirms the conclusions of the works
\cite{4}, \cite{6}, \cite{11} that there is a need to refine the notion of the order of a system which can not be considered only by the total number of differentiation. For fractional-order differential equations the number of terms is more important than the order of differentiation.

The results presented in this contribution give basis for
controlling chaotic fractional-order systems. An alternative
approximation of fractional-order derivative, stability
investigation, and also other chaotic
fractional-order system  will be used in further work.

\section*{Acknowledgements}
\noindent

This work was partially supported by grant VEGA 1/7098/20 from
the Slovak Grant Agency for Science.

\end{document}